%% file: epsl.tex
\documentclass[a4paper]{article}
\usepackage{psfig}
\def\bild#1{\def\thefigure{\arabic{figure}}%
   \refstepcounter{figure}\label{#1}~\thefigure}
\def\tabelle#1{\def\thetable{\arabic{table}}%
   \refstepcounter{table}\label{#1}~\thetable}
\def\eps{\varepsilon}

\usepackage{amssymb}
\usepackage{amsmath}

\begin{document}

\title{
Energy barriers in three-dimensional micromagnetic models and the physics of thermo-viscous magnetization in multidomain particles}

\author{Karl Fabian \\
	Geological Survey of Norway,\\
Leiv Eirikssons vei 39,\\
7491 Trondheim,\\
Norway,\\
karl.fabian@ngu.no
	\and
	Valera P. Shcherbakov  \\
	Geophysical Observatory 'Borok',\\  Yaroslavskaja Oblast, 151742,\\ Russia \\
	}

\maketitle

\begin{abstract}
A first principle micromagnetic and statistical
calculation of
viscous remanent magnetization (VRM)  in an ensemble of cubic magnetite
pseudo single-domain particles
is presented.
This is achieved by developing
a fast relaxation algorithm for finding optimal
transition paths between micromagnetic local energy minima.
It combines a nudged elastic band technique with action minimization.
Initial paths are obtained by repetitive minimizations of
modified energy functions.
For a cubic pseudo-single domain particle, 60 different local energy minima are
identified and all optimal energy barriers between them
are numerically calculated for zero external field.
The results allow to  estimate also the energy barriers in
in weak external fields which are necessary to
construct the  time dependent  transition  matrices which describe  the continuous
homogeneous Markov processes of VRM acquisition and decay.
By   spherical averaging the
remanence acquisition in an isotropic PSD ensemble was calculated
over all time scales.
The modelled particle ensemble shows a physically meaningful overshooting
during VRM  acquisition.
The results also explain why
VRM acquisition  in PSD particles can occur much faster than
VRM decay and therefore can explain  for findings of extremely
stable VRM in some paleomagnetic studies.
\end{abstract}

\section{Introduction}
\input{intro}

 \section{Action and path integrals}
\input{action}

\section{A modified relaxation method to determine transition paths from micromagnetic models}
\input{relax}
\section{Energy barriers in a cubic pseudo-single domain particle}
\input{results}
%\subsection{First principles calculation of magnetic viscosity}
\input{markov}
\section{Viscous magnetization}
\input{discu}
\section{Appendix}
\input{append}

\bibliography{kf,vrm,trm,hys}
\bibliographystyle{amsplain}

\end{document}

%% file: intro.tex
\subsection{Aim and outline of the article}

The geomagnetic field has been perpetually recorded by magnetic remanence
carriers in newly formed rocks throughout the Earth's history.
Therefore,
crustal rocks form a paleomagnetic archive which is accessible  through
 rock magnetic measurements.
 Yet, their interpretation requires a thorough understanding of
the physical processes occuring during  remanence acquisition.
Thermo-viscous magnetization of natural pseudo-single or multidomain
particles is   the most abundant
remanence  in paleomagnetism, although
more reliable single domain (SD) remanence carriers are preferred, and theoretical interpretation
is based on paradigms developed from SD theory.
The main aim of this article is to propose new computational and conceptional
methods to obtain a physical understanding of the remanence acquisition in
multidomain particles.
The second section will introduce a new technique to determine energy barriers in
micromagnetic models of inhomogenously magnetized particles.
Although this is technically challenging, it is an essential prerequisite
for a quantitative study of thermo-viscous magnetization.
In the third section we use the new computational method to calculate all energy barriers for a
three-dimensional model of metastable flower and vortex-states in a cubic magnetite particle.
The fourth section introduces the theoretical background for a statistical analysis  of viscous
remanence acquisition and decay in weak fields, based on the computed
energy barriers between the metastable states.
In the fifth section this theory is applied to the energy barriers calculated in section~3, and
the physical meaning of the model results obtained is discussed.

\subsection{Micromagnetic modeling}
Micromagnetic modeling is now a standard technique to determine
stable and metastable magnetization states in small ferro- and ferrimagnetic particles. It is used to calculate and analyze  magnetization structures
          in natural and synthetic magnetic nanoparticles.
          This size range is of special interest in rock magnetism
    where the reconstruction of the Earth magnetic field depends critically
    on the reliability of remanence carriers in natural rocks.
    The grain size distribution of these remanence carriers
    rarely is confined to the relatively small SD-size range.
     Accordingly, detailed knowledge of the
    physical mechanism of magnetization change in larger nanoparticles is
    needed to assess and evaluate the magnetic measurement results from natural
    materials.
Because of its importance for understanding remanence acquisition in natural rocks, rock magnetic studies were among the first to apply numerical micromagnetic models.
The first approach to estimate barriers between single-domain and two-domain states
used a one-dimensional model of magnetization change \cite{Enkin:88}.
When three-dimensional models were developed
to understand inhomogenous remanence states \cite{Williams:89}, it was immediately a main interest
to obtain energy barriers to model the acquisition of thermoremanence \cite{Enkin:94,Thomson:94,Winklhofer:97,Muxworthy:03}.
Knowing the energy barriers between
different magnetization states within a single particle
also leads to a  quantitative prediction  of magnetic viscosity and
magnetic stability of remanence information, even over geological time scales.
An important result of early micromagnetic calculations was that beyond the regime where exchange
    forces
    dominate, i.e. beyond length scales of several exchange lengths $\sqrt{A/K_d}$,
    there exist a multitude of local energy minima corresponding to meta-stable magnetization
    structures \cite{Williams:89,Fabian:96,Rave:98}.
In the context of thermo-viscous remanence, the most important property of meta-stable magnetization
    structures $m$ is their
    residence time $\tau(m)$. It denotes the expectation value of the
    time during which the system remains in
    state $m$, if it initially is in this state at time $t=0$.
The residence time $\tau(m)$ directly depends on the transition probabilities
    $p(m,m')$ between $m$ and all other LEM $m'$, which in turn depend upon the
    possible transition pathways.
To determine the transition probability $p(m,m')$ in very good approximation,
    it is sufficient to find the   most likely transition path between $m$ and  $m'$,
    which is the path with  the lowest energy barrier.
    This path  runs across the saddle-point with lowest energy of all
    which connect  $m$ and $m'$.
Therefore, the problem of finding the
    transition probabilities $p(m,m')$ is closely related to finding saddle-points in high-dimensional
    micromagnetic energy-landscapes.
    
\subsection{Statistical theory}
Given these transition probabilities, the geologically important mechanism of
thermoremanence acquisition can be described as a
 stochastic process of
magnetization change in a temperature dependent energy landscape.
Its transition matrix is related to $p(m,m')$, determined by the energy barriers between the possible
states \cite{Fabian:03b}.

%% file: action.tex
\subsection{Micromagnetic modeling}
  Berkov \cite{Berkov:98a,Berkov:98b,Berkov:98c}
developed a numerical method to evaluate   the distribution of energy barriers between
 metastable states in many-particle systems which determines the optimal path
between the two given metastable states by minimizing
the action in the Onsager–Machlup functional \cite{Onsager:53}
for the transition probability.
This method essentially performs a local saddle-point search in an high-dimensional
energy landscape.
Mathematically similar problems
exist in several disciplines of
physics and chemistry.
In the last years, several new methods to locate saddle-points
 have been developed in these fields \cite{Henkelman:00a,Henkelman:00b,Olsen:04}.
Based on such algorithms, an improved elastic band technique
for micromagnetics was presented by \cite{Dittrich:02}.

The main problem in energy-barrier computation is that micromagnetic structures $m$ are
described by many variables and accordingly
energy $E$ is a function of $m$.
Minimizing $E(m)$ requires sophisticated algorithms, but for
energy-barrier calculations it is even necessary to
determine saddle points in this high-dimensional energy landscape.

Several approaches are available for this task, but because saddle-point calculation
is equivalent to minimizing $(\nabla E(m))^2$ all rapidly converging
methods require second derivatives of $E$.
This however is rather to be avoided if the calculation should be performed
effectively.

The present study develops a combination of several of the above cited techniques to efficiently
calculate energy barriers in micromagnetic models.
\subsection{Action minimization}

Berkov \cite{Berkov:98a} introduced action minimization as a tool
for finding optimal transition paths in thermally driven
micromagnetic systems.
He discretized the time dependent action of the
the magnetic particle system and used a numerical quadrature
representation for direct numerical minimization.
This rigorous approach is complicated by its explicit dependence upon
transition time.
However, transition paths turn out to be geodesics of the energy surface
in the limit of infinite transition time, where
 energy barriers are lowest.

Dittrich~{\em et al.} \cite{Dittrich:02}
make use of this fact by directly searching for geodesic paths using a
modification of the nudged elastic band (NEB) algorithm
of Henkelman~{\em et al.}\cite{Henkelman:00a,Henkelman:00b}.

A problem of this algorithm is that it involves the numerical solution of a large
system of ordinary differential equations. Moreover, there is a tendency of
the NEB algorithm to produce spurious up-down-up movements
along the gradient (kinks) which cannot be completely removed in all cases.

Here, we combine both approaches by designing a
 path relaxation algorithm
similar to NEB, but constraint to decrease the action at each step.
The algorithm performs a fast gradient-like relaxation from an initial path
towards the optimal transition path. It
  detects and avoids the development of kinks, and does not involve
numerical solutions of differential equations.

The important problem of finding an
initial path which is likely to lie in the basin of attraction
of the optimal transition path under the proposed relaxation scheme
is also investigated.

\subsection{Geometric action}
Here we define the {\em geometric action} for a path $p$ as the minimal action for
any transition along this path.

In a general mechanic system the action of a transition process $x(t)$ from state
$x(0)=x_0$ into $x(t_{end})=x_1$ is defined
by
\begin{equation}\label{action-1}
    S(x(t)) ~:=~ \int \limits_0^{t_{end}} \langle \dot{x} + \nabla E, \dot{x} + \nabla E, \rangle\,dt.
\end{equation}
The probability that this transition process occurs depends monotonously on
$\exp(-S(x(t)))$.
In the next section we will be looking for the optimal transition path
in the energy landscape determined by $E$.
The quality of any given path $p$ is defined as its geometric action:
the action of the transition process $x_p(t)$
along   $p$ which minimizes (\ref{action-1}).

We start with a canonical parametrisation of $p$ by  arc length $s$
and try  to find  a reparametrisation $s(t)$ which minimizes $S$ along $p$.
For this optimal transition process we then have
\begin{equation}\label{action-2}
    S_{\rm min}(p)~=~S(x_p(t)) ~=~ \int \limits_0^{L(p)}
    \langle \frac{dx}{ds}\, v + \nabla E, \frac{dx}{ds}\, v  + \nabla E, \rangle\,\frac{ds}{v},
\end{equation}
where $L(p)$ is the arc length and $v(s)~=~ \frac{ds}{dt}(s)$ is the local velocity of the
optimal transition at arc length $s$.
Finding $s(t)$  thus is reduced to the  variational problem of finding
the function $v(s)$ which minimizes (\ref{action-2}).
The corresponding Euler-Lagrange equation is
\begin{equation}\label{Euler-1}
   \frac{d}{dv} \left( \frac{1}{v}\,
   \langle \frac{dx}{ds}\, v + \nabla E, \frac{dx}{ds}\, v  + \nabla E, \rangle \right) ~=~0.
\end{equation}
A short calculation confirms that it's solution is
\begin{equation}\label{Euler-2}
  v~=~ \|\dot{x}\|~=~ \|\nabla E\| \,\left\|\frac{dx}{ds}\right\|^{-1} ~=~\|\nabla E\|.
\end{equation}
The last equality uses the fact that for the arc length parametrisation
$\| {dx}/{ds}\|~=~1$. Inserting this result into (\ref{action-2}) yields
\begin{equation}\label{action-3}
    S_{\rm min}(p)  ~=~ 2\,\int \limits_0^{L(p)}
    \|\nabla E\|  \, \left\|\frac{dx}{ds}\right\| + \left\langle\frac{dx}{ds},\nabla E  \right\rangle~ds
    ~=~ 2\,\int \limits_0^{L(p)}
    \|\nabla E\|    + \left\langle\frac{dx}{ds},\nabla E  \right\rangle~ds.
\end{equation}
This integral can be simplified further by noting that
\begin{equation}\label{action-4}
 \int \limits_0^{L(p)} \left\langle\frac{dx}{ds},\nabla E  \right\rangle~ds ~=~
 \int \limits_{E(x_0)}^{E(x_1)}  dE  ~=~ E(x_1) -E(x_0)~=:~\Delta E.
\end{equation}
Accordingly, one obtains the geometric action of $p$
as
\begin{equation}\label{action-5}
    S_{\rm min}(p)  ~=~  2\,\Delta E~+~2\,\int \limits_0^{L(p)}
    \|\nabla E\| \,ds.
\end{equation}

\subsection{Finding the optimal path by variation of the geometric action}
It is possible to find the Euler-Lagrange equations for the
optimal path by variation of the geometric action
$S_{\rm min}(p)$ with respect to $x$.

To this end we reparametrize (\ref{action-5}) by $w(s)=s/L(p)$ and obtain
\begin{equation}\label{action-6}
     S_{\rm min}(p)  ~=~  2\,\Delta E~+~ 2\,\int \limits_0^{1}
    \|\nabla E\| \, \left\|\frac{dx}{dw}\right\|  \,dw.
\end{equation}
The    Euler-Lagrange equation
of the variational problem $ \delta S_{\rm min}(p) ~=~0$
after some simplification
has the form
\begin{equation}\label{EL-2}
\frac{d^2x}{ds^2}~=~  \nabla\, \log\|\nabla E\|.
\end{equation}
The details of the calculation are given in the appendix.

\subsection{The optimal path is a geodesic}
A path along an energy surface which fulfills
 \begin{equation}\label{TP-1}
\dot{x}~=~  \pm  \nabla E
  \end{equation}
is a geodesic.
In the one-dimensional case (\ref{Euler-2}) directly  implies that
the optimal transition path is a geodesic.
In the multidimensional case this
not simply  follows from (\ref{Euler-2})
which is valid for any
geometric transition path.
Yet, by applying the Cauchy inequality to
(\ref{action-6}) one obtains  for the optimal  path
$p$
\begin{equation}\label{TP-3}
     S_{\rm min}(p)  ~\geq~  2\,\Delta E~+~ 2\,\int \limits_0^{1}
    \left|\left\langle \nabla E ,\, \frac{dx}{dw}\right\rangle\right|  \,dw.
\end{equation}
The integration interval $[0,1]$ can be divided into
finitely many parts $[w_k,w_{k+1}]$
with alternating constant sign of $ \langle \nabla E ,\, dx/dw \rangle$.
Accordingly, $\nabla E(x(w_{k}))~=~0$ and
\begin{equation}\label{TP-4}
     S_{\rm min}(p)  ~\geq~  2\,\Delta E~+~ 2\,\sum \limits_{k=0}^{K}
    \left| \, E(x(w_{k+1}))- E(x(w_{k})) \right|.
\end{equation}
Here the right hand side is a lower limit of $ S_{\rm min}$.
Accordingly, a geodesic  which fulfills (\ref{TP-1}) achieves equality
in (\ref{TP-3}). It therefore coincides with the least action path between
the prescribed endpoints.
Equality in (\ref{TP-4}) means that the least action depends only upon the energies at
the traversed  critical points.

\subsection{Morse theory}
Topologically different critical points in multi-dimensions are distinguished
by their {\em Morse index}, which is defined as the dimension $n_-$
of the sub-manifold on which the Hessian is negative definite.
Intuitively, the Morse index of the highest saddle point along the optimal transition path
should not be too large. This is  because , apart from singular cases,
  the action is minimized along an only one-dimensional manifold,
   i.e. an isolated path parallel to the gradient which connects initial and final minima.
Therefore, if  at the  highest saddle point the Morse index is  $n_- > 1$,
the other $n_- - 1$ descending directions should lead into different LEM states.
The choice for these LEM states should not be too large whenever the initial and final minima
are close to the global  energy minimum.

On the other hand, Morse theory \cite{Milnor:63M} implies that the total number of saddle-points
will be huge in realistic micromagnetic calculations.
For a two-dimensional  surface  with Euler characteristic
$\chi_{\rm Euler}$,
the numbers
$N_{\rm min}$ of minima,
 $N_{\rm max}$ of maxima, and
 $N_{\rm saddle}$ of  saddle points
 are connected by the relation
\begin{equation}\label{crit}
 N_{\rm min} + N_{\rm max} -N_{\rm saddle} ~=~ \chi_{\rm Euler}.
 \end{equation}
For a sphere is $\chi_{\rm Euler}=2$.
The generalization of (\ref{crit}) to a finite-dimensional
compact manifold is the {\em Morse relation}
 \begin{equation}\label{Euler-N}
 \sum \limits_{k=0}^{N} (-1)^k N_{k} ~=~ \chi_{\rm Euler}.
 \end{equation}
Here $N_{k}$  is the number of critical points
with Morse index  $k$, i.e.
where the negative definite sub-manifold has dimension $n_- = k$.
In case of the micromagnetic energy depending on $N$ magnetization directions
  the manifold is a direct product of $N$ two-dimensional
spheres, therefore   $ \chi_{\rm Euler} = 2^N$.
On this $2 N$-dimensional manifold we get from (\ref{Euler-N})
 \begin{equation}\label{crit-N}
 N_{\rm min} + N_{\rm max}+ N_{{\rm even}} -N_{ {\rm odd}} ~=~ 2^N,
 \end{equation}
where $N_{ {\rm even}}$ and $N_{{\rm odd}}$ are the number of
true saddle points
with  even or odd Morse index, respectively.

If this manifold describes  an ensemble of
 interacting SD grains
 each grain has at least two critical points,
 minima or maxima,  in zero external field.
 If   interaction is weak, the total energy of the system   inherits almost
 all these minima and maxima as saddle-points.
  Thus, it is understandable that the total number of critical points exceeds even a huge figure
  like $2^N$.

  The situation is different if the $2 N$-dimensional manifold
  describes an exchange coupled grain with inhomogeneous magnetization structure,
  like in   most micromagnetic applications.
  In this case, it turns out that only a limited number of   minima and maxima,
  like flower  or  vortex states, exist.
     According to (\ref{crit-N}), there must appear an enormous number
     of true saddle points  with even Morse index -- i.e. not one-dimensional lines.

     Thus, inevitably there exists a large number of paths,
     with different action $S$,
     which  connect the local minima and complicate the  search of the lowest action path.
     Another circumstance causing difficulties to find path lines by numerical means,
     is that the field lines passing through the saddle points,
     form sub-manifolds of zero measure.
     In other words, the saddle points are unstable in the sense,
     that almost all field lines in their vicinity are deflected aside
     (except for those which go directly through the saddle point).

\subsection{Euler-Lagrange path relaxation}
To numerically calculate the optimal geometric transition path
one would like to start from an arbitrarily chosen initial
path $p_0$ which then is iteratively improved
by some updating scheme.
An heuristic procedure is known as 'nudged elastic band' technique.
Here we derive an optimal updating scheme which corresponds to
a gradient minimization of minimal path action.
To this end we follow the standard derivation of the Euler-Lagrange
equation to prove that the change $\delta S_{\rm min}$ in action
due to a variation $\delta x$ in the path is given by
\begin{equation}\label{EL-4}
\delta S_{\rm min}~=~
2\,\int \limits_0^L \left( \nabla \|\nabla E\| - \|\nabla E\| \, \frac{d^2x}{ds^2}
\right)
\,\delta x\,ds.
\end{equation}
The direction of maximal increase in $S_{\rm min}$ now is a function
$\delta^* x \in L_2([0,L])$ with $\|\delta^* x\|_2~=~1$ which maximizes
(\ref{EL-4}). By Cauchy's inequality
\begin{equation}\label{optstep}
\delta^* x~=~
\frac{   \nabla \|\nabla E\| - \|\nabla E\| \, \frac{d^2x}{ds^2}
 }{
\left\| \nabla \|\nabla E\| - \|\nabla E\| \, \frac{d^2x}{ds^2}
\right\|_2}.
\end{equation}
The optimal Euler-Lagrange relaxation towards a minimal
 action path accordingly is given by
\begin{equation}\label{EL-update}
    x_{n}~:=~ x_{n-1} - \alpha\,\delta^* x_{n-1},
\end{equation}
for sufficiently small  positive step size $\alpha$.
Two points are noteworthy. First, the updating depends only on $\|\nabla E\|$ and not
explicitly on $E$, and second, the term $d^2x/ds^2$ leads to reduction of local curvature
in regions where $\|\nabla E\|>0$. This prevents the updating scheme from producing
 kinks. Only at points with  $\|\nabla E\|=0$ the final
 solution may not be differentiable.

\subsection{Relation between action and path relaxation}
There exists a heuristic scheme to find a path between two minima
 across a saddle-point
in multidimensional energy landscapes, which
makes use of the fact that there is such a path which
 always runs along the energy gradient.

Starting from an arbitrary initial path $x_0(s)$ it is thus attempted to
adjust the path in the direction along the negative energy gradient, but perpendicular to the
local tangent.
This relaxation scheme can be analytically described as the solution of a boundary
value problem for a system of nonlinear partial differential equations
\begin{equation}\label{pde-1}
    \frac{\partial x}{\partial u}(u,s)~=~
    - \nabla E + \frac{\langle \frac{\partial x}{\partial s}, \nabla E \rangle}{\| \frac{\partial x}{\partial s}\|^2}
    \, \frac{\partial x}{\partial s},
\end{equation}
with the boundary conditions
\begin{equation}\label{bnd-1}
 ~~~x(0,s)~=~x_0(s),~x(u,0)~=~x_0(0),~x(u,L)~=~x_0(L).
\end{equation}

%% file: relax.tex
\subsection{Definitions}
A micromagnetic structure $m$ is determined by
$K$ magnetization vectors on a spacial grid over the particle.
Each magnetization vector is a unit vector determined by two
polar angles $\theta, \phi$.
The distance $d(m_1,m_2)$ between two magnetization structures is defined by
\begin{equation}\label{dist}
   d(m_1,m_2)~:=~ \left[ \frac{1}{V} \, \int \limits_V \arccos^2 ( m_1(r) \cdot m_2(r) ) \,dV\right]^{1/2}.
\end{equation}
The local direction vector from $m_1$ to $m_2$ is the gradient
\begin{equation}\label{dir}
  v(m_1,m_2)~:=~ -\nabla  d(\,.\,,m_2) (m_1).
\end{equation}

If two magnetization structures $m_0$ and $m_1$ contain no
opposite directions, i.e. for all $r\in V$ we have $ m_0(r) \neq -m_1(r)$,
then it is possible to linearly interpolate between $m_0$ and $m_1$ by
defining $ m_t(m_0,m_1) (r) $ as the intermediate vector
on the smaller  great circle segment connecting $m_0(r)$ and $m_1(r)$ which
has angular distance $t \arccos(m_0(r) \cdot m_1(r))$ from $m_0(r)$.

By minimization of $E(m)$
using gradient information $\nabla E (m)$
an initial minimum $m_A$ and a final minimum $m_B$
are found.

\subsection{Outline of the relaxation procedure}
In summary, the results of the previous section show that the transition probability
between $m_A$ and   $m_B$ is in very good approximation determined by the 
minimal energy barrier between them.
This barrier is achieved along some optimal geometrical least action path 
$m(s)$ with $m(0)~=~m_A$ and $m(1)~=~m_B$ which also represents the most 
likely transition path.
The state of maximal energy along this path is a saddle point of the 
total energy function $E(m)$, and the least action path is  everywhere parallel to 
$\nabla E(m)$.
To find this optimal transition path, we propose a relaxation method 
which combines the advantages of the NEB technique of \cite{Dittrich:02}
with the additional action minimization of \cite{Berkov:98b}.

Two techniques are required for finding the optimal path by means of iterative 
relaxation: 
\begin{enumerate}
  \item  An updating scheme which determines an improved transition path $m^{k+1}(s)$
  from a previous path $m^k(s)$.
  \item A method of finding an initial transition path $m^0(s)$ within the basin of attraction of 
   the optimal path.
\end{enumerate}

The here proposed updating scheme starts from an
initial path $m^0(s)$ which is 
determined  by interpolating  $N$ intermediate states
$m^0(s_j)$ which correspond to the magnetization structures at the 
$s$-coordinates $0=s_1<s_2<\ldots<s_N=1$ for $j=1,\ldots, N$.
For the interpolation to be well-defined,  $N$ is required to be large enough 
to ensure that neighboring structures $m^k(s_j)$ and $m^k(s_{j+1})$ never contain opposite 
magnetization vectors. 

Similar to the NEB method, in step $k$ the path $m^k(s)$ is changed according to
\begin{equation}\label{relax}
    m^{k+1}(s) ~:=~  m^{k}(s) - \alpha_k \, \left[ \nabla E(m^k(s))-
          \left( \nabla E(m^k(s)) \cdot t^k(s)  \right)\,t^k(s)\right],
\end{equation}
where $t^k(s)$ is the tangent vector to the path $m^k(s)$  at $s$ and
$\alpha_k>0$ is a real number.
This updating scheme moves the path downward along the part of the energy gradient which
is perpendicular to path itself. This algorithm 
converges to a path which is almost everywhere 
parallel to $\nabla E$.
However,  (\ref{relax}) describes  not a true gradient 
descent for the action, and the final path may not achieve minimal action
  due to the formation of kinks during the minimization 
\cite{Henkelman:00a,Henkelman:00b,Dittrich:02}.

The here proposed method differs from previous NEB techniques  
in two details:
First,  
$\alpha_k$ is chosen such that $S( m^{k+1})< S( m^{k})$. This ensures that 
the action decreases in each step.
Second, $\alpha_k$ is dynamically adapted to achieve rapid convergence.
The following procedure to choose  $\alpha_k$  
fulfills both aims.
 
\subsubsection*{Adaption of $\alpha_k$}
Starting with the  initial value $\alpha_0=1$,
it is evaluated after each step whether the action of the updated path is decreased, i.e.
$ S( m^{k+1}(s)) < S( m^{k}(s))$. 
While this is true, the new value
$\alpha_{k+1}=\alpha_k$ is kept constant, but only for at most five steps. 
In this phase (\ref{relax}) performs a  quasi-gradient descent 'creeping' towards 
the optimal path. 

Afterwards,
if $S$ still decreases, we set $\alpha_{k+1}=2\,\alpha_k$ 
in each following step until
some $ S( m^{k+1}(s)) > S( m^{k}(s))$.
This phase can be interpreted as an 'accelerated steepest descent'.

If at any time $ S( m^{k+1}(s)) > S( m^{k}(s))$, 
the path $ m^{k+1}(s)$ is rejected and (\ref{relax})
is evaluated for 
a new  $\alpha_{k+1}$-value  of
$\alpha_{k+1}=1/4\,\alpha_k$.
This behavior avoids 'overshooting' of the gradient descent and

All our tests show that this iterative adaption of $\alpha_k$ 
leads to a  much faster convergence than
choosing any fix value $\alpha_k=\alpha$.
 
By comparing the achieved action $S( m^{k}(s))$ to $ S_{\rm min}$ from
(\ref{TP-4})
during the relaxation, it is possible to detect  the 
formation of kinks and to decide when the minimization succeeds.
This action criterion is better than testing whether 
the final path is parallel to $\nabla E$, since
the latter is also true for paths with kinks.
 
\subsection{Determination of the initial path}
Since the relaxation scheme works similar to a gradient
minimization algorithm, it adjusts to the next local optimum of the
action function.
Therefore the choice of the initial path is crucial for obtaining the globally
optimal path.
Here we propose a  method to find a good initial path $m^0(s)$ by
a sequence of minimizations of modified energy functions.

For parameters $\mu, \beta, \eps$ and $\Delta$ we consider the
modified energy function
\begin{equation}\label{modEn}
   E^\ast_\Delta(m) ~=~
         E(m) + \mu \,[ d(m, m_A) -\Delta]^2 + \frac{\beta}{d(m,m_B)+\eps}.
\end{equation}
The parameters  $\mu, \beta, \eps$ are chosen such that
for $\Delta=d(m_B, m_A) $ the final state $m_B$ is a unique
optimum of $E^\ast_\Delta(m)$, while for $m$ with  $d(m , m_A) \ll d(m_B, m_A)$ the last term of
$E^\ast_\Delta(m)$ is small in comparison to $E(m)$.
The value of $\mu$ should be large enough to ensure that the distance between
$m_A$ and the minimum of
$E^\ast_\Delta$ is indeed close to $\Delta$.

Now a sequence $m^0_j,~j=1,\ldots, J$ of magnetization structures is iteratively
obtained by setting $m^0_0~=~m_A$ and
$m^0_{j}$ to the result of minimizing $E^\ast_{\Delta_j} (m)$, where
$\Delta_j~=~j/J\,d(m_B, m_A)$.

Interpolating between the  structures  $m^0_j$ determines an
initial path  which (1) starts in $m_A$ and ends in $m_B$,
(2) has relatively equally spaced
intermediate states $m^0_{j}$, and (3) prefers
 intermediate states $m^0_{j}$ at distance $\Delta_j$ with low energy $E(m)$.

%% file: results.tex
\subsection{Material constants}
The numerical calculations have been performed in terms of
the  reduced material parameters $Q$ and $\lambda$ \cite{Hubert:98M}.
The magnetic hardness $Q$, in the case of cubic magneto-crystalline anisotropy, is the numerical ratio of
 $Q=K_1/K_d$. Here $K_d$ is the characteristic magnetostatic self-energy, which in terms of the
 saturation magnetization $M_s$ is defined as
 $K_d= 1/2 \mu_0 M_s^2$. The exchange length $\lambda= \sqrt{A/K_d}$ determines the characteristic length scale
 above which magnetostatic self-energy is able to overcome exchange coupling, represented
 by the exchange constant $A$.
For magnetite these
material constants are $M_s=480$~kA/m, $A=1.32\cdot10^{-11}$~J/m, and
$K_1=-1.25\cdot10^4$~J/m$^3$.
Accordingly,
$K_d=145$~kJ/m$^3$,  $\lambda_{ex}=9.55$~nm and
magnetic hardness is $Q=-0.0863$. Unless otherwise stated, all results in the following
subsections are obtained for the magnetite-like value $Q=-0.1$. Lengths $\lambda$ are given
in units of $\lambda_{ex}$.
\subsection{Potential barriers in cubic particles}
~\\[3mm]%
\par
\vbox{ \centerline{\hbox{%
\begin{tabular}{ccccc}
  \hline
  $\lambda$&Initial state& Initial energy $E$ & Intermediate state& Barrier $\eps$\\
   \hline
  \hline
 4.0 & F-111& 0.286404 & F-110 & 0.007825 \\
 4.5 & F-111& 0.28308 & V-011 & 0.005417 \\
 4.5 & F-111& 0.28308 & F-011 & 0.008172 \\
 4.5 & F-111& 0.28308 & V-001 & 0.011041 \\
  \hline
\end{tabular}
}} \footnotesize
%  \begin{center}
{\bf Table \tabelle{trans-barr-1} :}
Energy barriers for numerically found optimal transitions between
listed initial and final states for $Q=-0.1$ and
different values of $\lambda$.
%  \end{center}
\normalsize }
~\\[0.2cm]%

If to compare the barriers   with the SD case, where $\eps    = 1/120 = 0.008333$,
the actual barriers are less, especially for $\lambda =4.5$.
But it must be kept in mind that for $\lambda =4.0$ the transition is quasi-SD as the
saddle point is F-110 type and almost homogeneous. For  $\lambda =4.5$, however,
the minimum energy saddle point is V-110 type,
and development of vortex evidently considerably reduces the barrier.
~\\[3mm]%
\par
\vbox{ \centerline{\hbox{ \psfig{figure=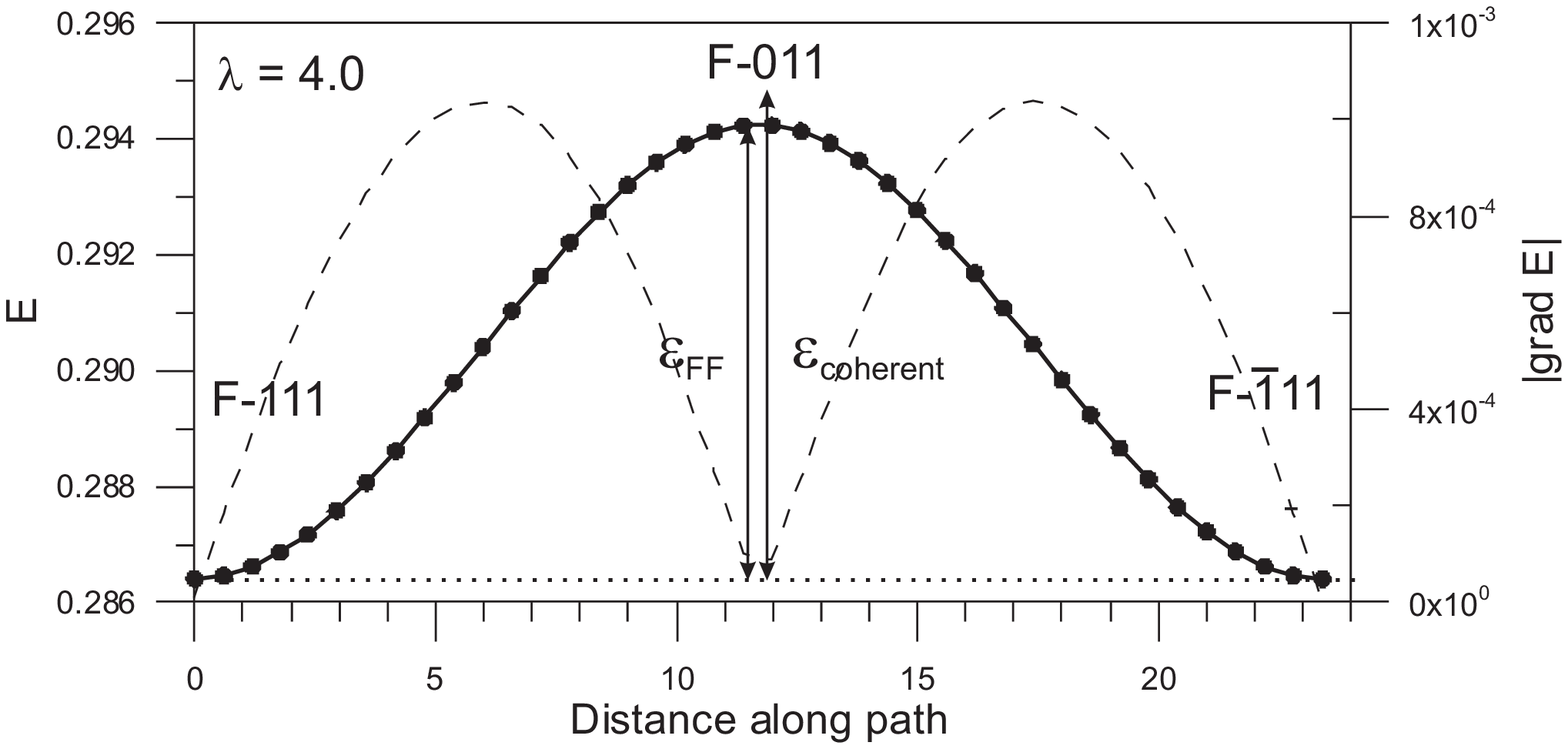,width=140mm}
}} \footnotesize
%  \begin{center}
{\bf Figure \bild{lam4-barr} :}
Energy variation across the optimal transition from a F-111 flower state to a
F-\={1}11 vortex state at $\lambda = 4.0$. Each circle represents an intermediate
 magnetization state used for the calculation. The dashed line corresponds to the
 absolute value of the energy gradient along the transition path. The maximum energy state along the
 path is the F-011 flower state. The three-dimensional micromagnetic calculation ($\eps_{FF}$)leads only to
 a minor decrease of the energy barrier with respect to coherent rotation ($\eps_{coherent}$).
This improvement results from the small spin deflections close to the particle surface.
%  \end{center}
\normalsize }
~\\[0.2cm]%

For  $\lambda =5.0$ the situation is most complex as there are a
number of competitive LEM-states  with similar energy.
V001 with E = 0.27079, V110 with E = 0. 276931, V111 with E = 0.277701,
F111 with E = 0.279527.
As the result, there are no direct jumps between topologically identical
states like V-001 and V-010.
Such transitions were the only ones  observed for lambda $\lambda =4.0$ and $\lambda =4.5$.
For $\lambda =5.0$  the state V-001 can change
  to V-010 only indirectly, either through V-111, or V-011, or F-111.
 Thus, the diagonal elements of the matrix below are empty.
~\\[3mm]%
\par
\vbox{ \centerline{\hbox{%
\begin{tabular}{ccccc}
  \hline
  &V-001&V-110&V-111&F-111\\
  \hline\hline
  V-001&0&0.000123&0.001389 & 0.00349\\
 V-110&0.006263&0&0.000598& 0.002925\\
 V-111 &   0.008297 & 0.001368&0   &     0.003836\\
 F-111  &  0.012117   & 0.005521   & 0.005665  &0\\
  \hline
\end{tabular}
}} \footnotesize
%  \begin{center}
{\bf Table \tabelle{trans-barr-2} :}
Energy barriers for optimal transitions between
a reduced set of initial and final states for $Q=-0.1$ and
  $\lambda=5.0$.
  Transitions with the same  initial state are listed in the same column.
Transitions with the same  final state are listed in the same line.
The corresponding complete set of 60 LEM states is obtained by taking into
account cubic symmetry leading to 8 states of class F-111 or V-111,
6 states of class V-001, and 12 states of class V-110.
In addition, all vortex states exist in two varieties of different helicity.
%  \end{center}
\normalsize }
~\\[0.2cm]%

~\\[3mm]%
\par
\vbox{ \centerline{\hbox{ \psfig{figure=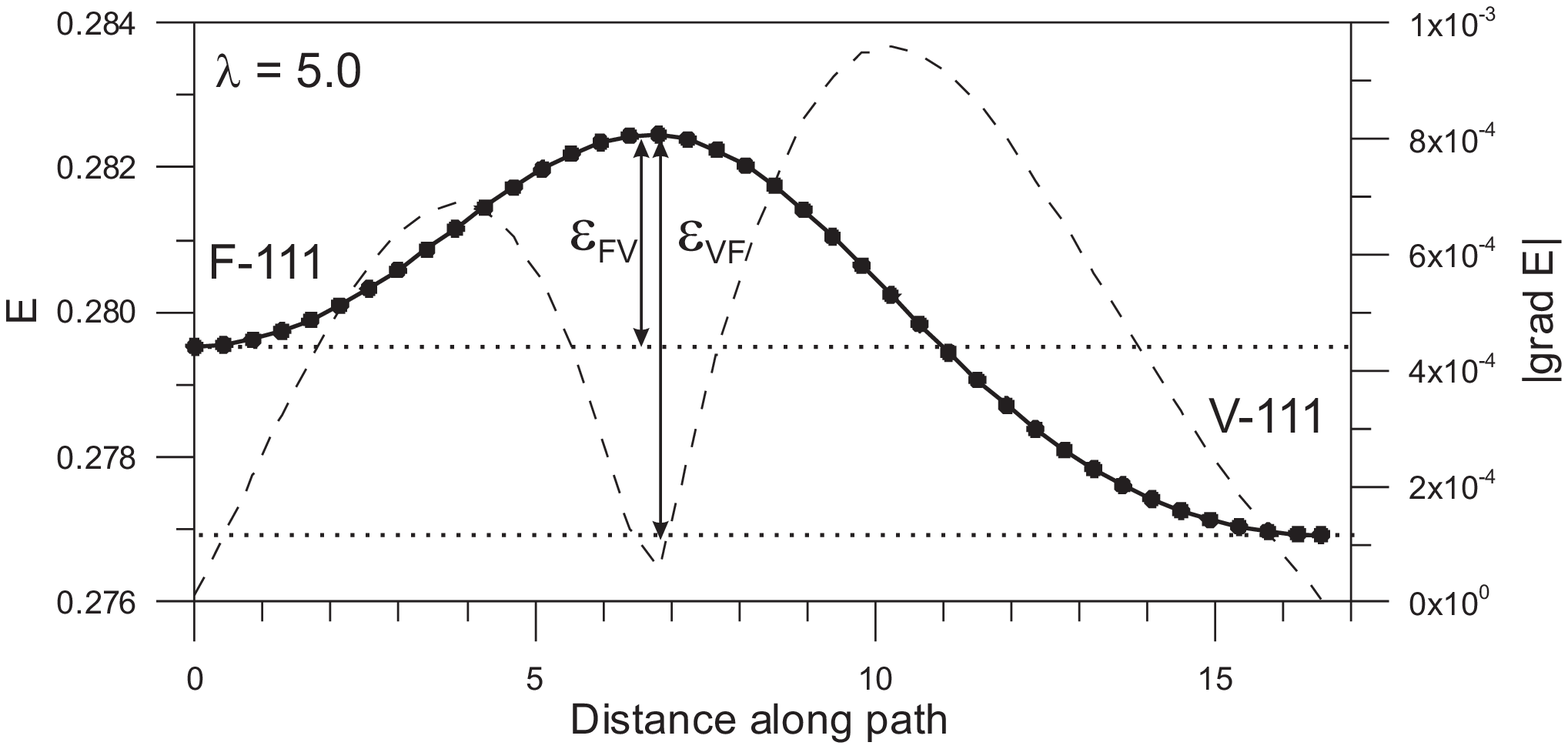,width=140mm}
}} \footnotesize
%  \begin{center}
{\bf Figure \bild{lam5-barr} :}
Energy variation across the optimal transition from a F-111 flower state to a
V-111 vortex state at $\lambda = 5.0$. Each circle represents an intermediate
 magnetization state used for the calculation. The dashed line corresponds to the
 absolute value of the energy gradient along the transition path. Because the energies of
 F-111 and V-111 are different, also the energy barriers for a transition from
 F-111 to V-111 ($\eps_{FV}$) and from V-111 to F-111 ($\eps_{VF}$) differ.
%  \end{center}
\normalsize }
~\\[0.2cm]%

For  $\lambda \geq 6.0$ the situation becomes simple again as now the only stable states are
of type V-001.
~\\[3mm]%
\par
\vbox{ \centerline{\hbox{%
\begin{tabular}{ccccc}
  \hline
  $\lambda$&Initial state&   Intermediate state& Barrier $\eps$\\
   \hline
  \hline
 6.0 & V-100&   V-111 & 0.009486  \\
 7.0 & V-100&  V-111 & 0.006372 \\
 8.0 & V-100&   V-111 & 0.005876  \\
  \hline
\end{tabular}
}} \footnotesize
%  \begin{center}
{\bf Table \tabelle{trans-barr-3} :}
Energy barriers for optimal transitions between
vortex states for $Q=-0.1$ and
  $\lambda=6.0-8.0$.
While V-111 is a marginally stable LEM at $\lambda=6.0$,
it is unstable for $\lambda=7.0,8.0$.
Types of V-100 vortex states are the global energy minima at these grain sizes
and the energy barrier refers to a symmetric transition between
two adjacent states of this type, e.g. V-100 to V-010.
%  \end{center}
\normalsize }
~\\[0.2cm]%

~\\[3mm]%
\par
\vbox{ \centerline{\hbox{ \psfig{figure=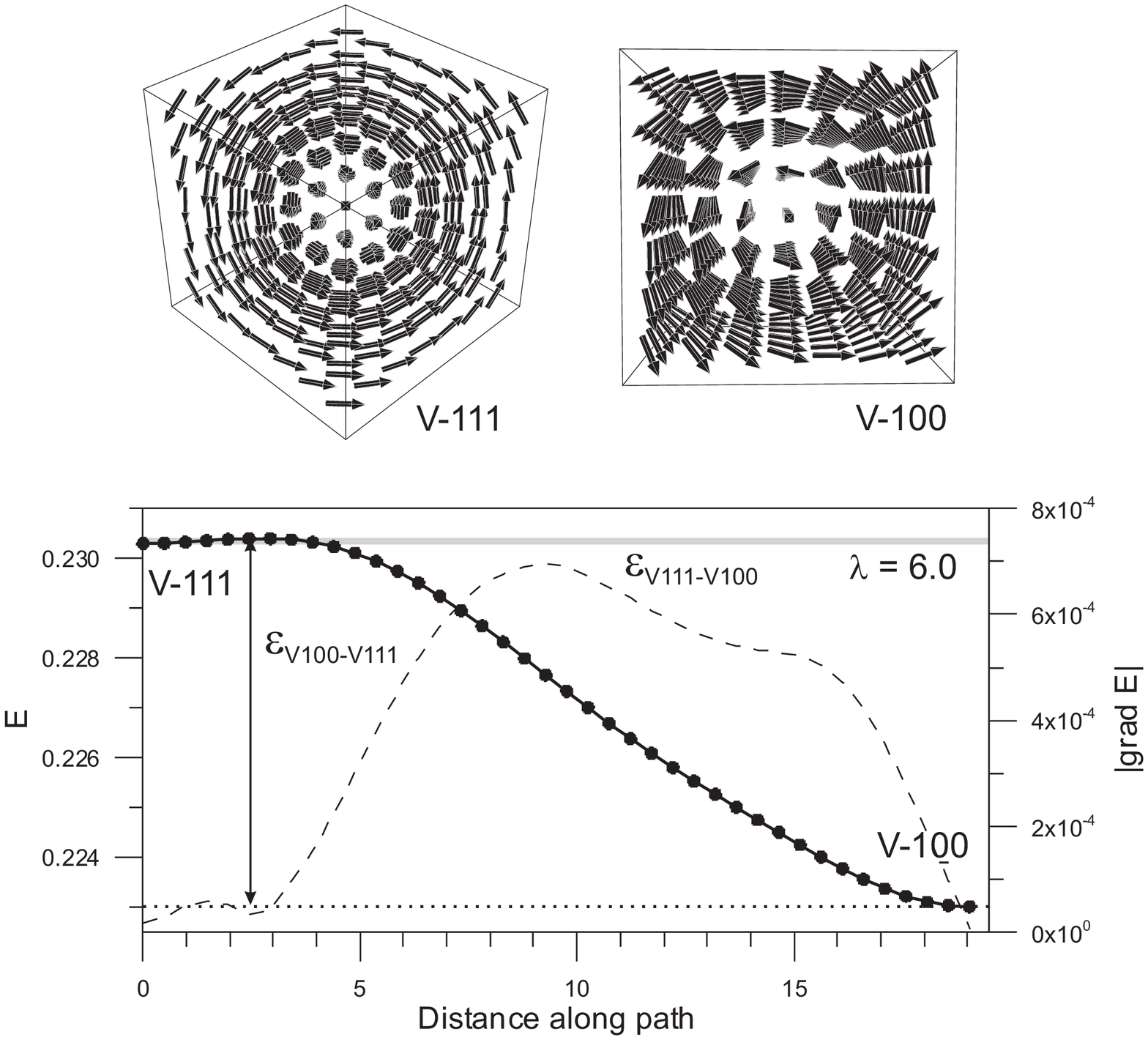,width=140mm}
}} \footnotesize
%  \begin{center}
{\bf Figure \bild{lam6-barr} :}
Energy variation across the  transition from a V-111 flower state to a
V-100 vortex state at $\lambda = 6.0$.
The top figures show the magnetization structure of the V-111 state as seen along the $\langle111\rangle$-direction
and the V-100 vortex as seen along the $\langle100\rangle$-direction.
In the bottom diagram again each circle represents an intermediate
 magnetization state used for the calculation. The dashed line corresponds to the
 absolute value of the energy gradient along the transition path.
 The tiny energy barrier $\eps_{V111-V100}$ in relation to $\eps_{V100-V111}$
  indicates that the V-111 state is a very unstable LEM
 as compared to V-100. However, it is important for the transition between
 the more stable vortex states. E.g. the optimal transition
 from  V-001 to V-100 is a
 combination of the symmetric vortex rotations V-001 to V-111, and the shown
 transition  V111 to V100.
%  \end{center}
\normalsize }
~\\[0.2cm]%

Consider now the thermo-activation barrier $E_b = \eps \,\lambda^3$
for the most stable LEM states:
  F-111 for   $\lambda \leq 4.5$, and V-001 for  $\lambda \geq 5.0$.
~\\[3mm]%
\par
\vbox{ \centerline{\hbox{%
 \begin{tabular}{cccc}
    \hline
    $\lambda$& $\eps$ & $E_b$& $m$\\
      \hline
  \hline
4.0&  0.007825  &  0.5008 & 0.991297\\
4.5& 0.005417  &  0.49362& 0.986527\\
5.0  & 0.006263  &  0.782875  &  0.705928\\
6.0  & 0.009486 &   2.048976 &   0.509672\\
7.0  & 0.006372  &  2.185596 &   0.310451\\
8.0  & 0.005876 &    3.008517 &\\
    \hline
  \end{tabular}
}} \footnotesize
%  \begin{center}
{\bf Table \tabelle{trans-barr-4} :}
Minimal energy density $\eps$ or
absolute energy $E_b$ necessary
to leave the global energy minima for different
values of $\lambda$.
$m$ is the reduced magnetization at the global minimum.
%  \end{center}
\normalsize }
~\\[0.2cm]%
~\\[3mm]%
\par
\vbox{ \centerline{\hbox{ \psfig{figure=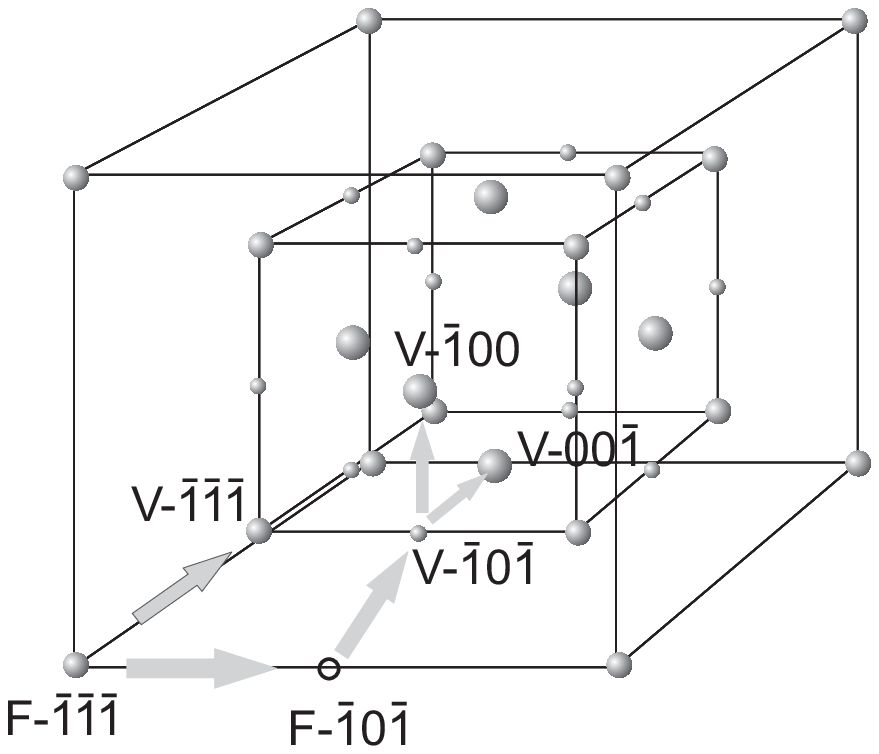,width=120mm}
}} \footnotesize
%  \begin{center}
{\bf Figure \bild{state-scheme} :}
Schematic representation of transition paths in magnetization space for a
cubic particle with $\lambda = 5.0$.
Each sphere corresponds to the magnetization of an LEM state. The cubic structure
reflects the cubic symmetry of the particle.
Grey arrows indicate two of the many possible transition paths (or decay modes):
(1) A direct decay form F-$\bar{1}\bar{1}\bar{1}$ to V-$\bar{1}\bar{1}\bar{1}$;
(2) Indirect decay from F-$\bar{1}\bar{1}\bar{1}$ over an instable intermediate
F-$\bar{1}0\bar{1}$ state into V-$\bar{1}0\bar{1}$ and
further to either V-$\bar{1}00 or V-00\bar{1}$.
Note, that each sphere in principle represents two vortex states of
inverse helicity (R and L). However, a transitions between
vortex states of different helicity
have large energy barriers and can be neglected.
%  \end{center}
\normalsize }
~\\[0.2cm]%

%
%
%Statelist:\\
%\footnotesize
%F$\bar{1}\bar{1}\bar{1}$, F$\bar{1}\bar{1}1$, F$\bar{1}1\bar{1}$,
%F$\bar{1}11$, F$1\bar{1}\bar{1}$, F$1\bar{1}1$, F$11\bar{1}$, F$111$,\\[3mm]
%V$_{-}00\bar{1}$, V$_{+}00\bar{1}$, V$_{-}001$, V$_{+}001$, V$_{-}0\bar{1}0$,
%V$_{+}0\bar{1}0$, V$_{-}010$, V$_{+}010$,\\[3mm]
% V$_{-}0\bar{1}\bar{1}$,
%V$_{+}0\bar{1}\bar{1}$, V$_{-}0\bar{1}1$, V$_{+}0\bar{1}1$, V$_{-}01\bar{1}$,
%V$_{+}01\bar{1}$, V$_{-}011$, V$_{+}011$,\\[3mm]
% V$_{-}\bar{1}00$, V$_{+}\bar{1}00$,
%V$_{-}100$, V$_{+}100$,\\[3mm]
% V$_{-}\bar{1}0\bar{1}$, V$_{+}\bar{1}0\bar{1}$,
%V$_{-}\bar{1}01$, V$_{+}\bar{1}01$, V$_{-}10\bar{1}$, V$_{+}10\bar{1}$,
%V$_{-}101$, V$_{+}101$,\\
% V$_{-}\bar{1}\bar{1}0$, V$_{+}\bar{1}\bar{1}0$,
%V$_{-}\bar{1}10$, V$_{+}\bar{1}10$, V$_{-}1\bar{1}0$, V$_{+}1\bar{1}0$,
%V$_{-}110$, V$_{+}110$,\\[3mm]
% V$_{-}\bar{1}\bar{1}\bar{1}$,
%V$_{+}\bar{1}\bar{1}\bar{1}$, V$_{-}\bar{1}\bar{1}1$, V$_{+}\bar{1}\bar{1}1$,
%V$_{-}\bar{1}1\bar{1}$, V$_{+}\bar{1}1\bar{1}$, V$_{-}\bar{1}11$,
%V$_{+}\bar{1}11$,\\
% V$_{-}1\bar{1}\bar{1}$, V$_{+}1\bar{1}\bar{1}$,
%V$_{-}1\bar{1}1$, V$_{+}1\bar{1}1$, V$_{-}11\bar{1}$, V$_{+}11\bar{1}$,
%V$_{-}111$, V$_{+}111$\\
%\normalsize 

%% file: markov.tex
\subsection{Statistical theory of MD VRM}
The first application of the above presented calculation of energy barriers is a complete
description of low-field viscous magnetization processes in a micromagnetically modelled
cubic particle.
Knowing the optimal transition paths between all  LEM structures $S_i$
of the investigated particle
allows for calculating the zero field temporal isothermal transition matrix $M(\Delta t)$,
which describes the continuous homogeneous Markov process of random thermally activated transitions
between all possible states:
\begin{equation}\label{transmat}
   M (\Delta t)~\equiv~\mathbb{P}\left[{ S(t)= S_j  \wedge S(t+\Delta t)= S_i }\right]~=~
   \,\exp\left[\mu \, \Delta t \right].
\end{equation}
Here the matrix elements $\mu_{ij}$ of the infinitesimal generator of the semigroup  $M(t)$ are given by
the relative outflow from $S_j$ to $S_i$ for $i\neq j$. The relative inflow from all other
states determines the diagonal element $\mu_{ii}$.
\begin{eqnarray}
% \nonumber to remove numbering (before each equation)
  \mu_{ij} &=& -\frac{\Delta E_{ij}}{k_B T\,\tau_0} ~~\mbox{for}~~i\neq j\\
  \mu_{ii}&=& -\sum \limits_{i\neq j}  \mu_{ji}
\end{eqnarray}
Once, the $ \mu_{ij}$ have been calculated, it is easily possible to determine the viscous decay
of any initial probability distribution
$\rho_i^0~\equiv~\mathbb{P}\left[{ S_0=  S_i }\right]$
by   multiplication with the time evolution matrix exponential
\begin{equation}\label{probdens}
   \rho( t) ~=~\exp\left[\mu \,  t \right]\, \rho_i^0
\end{equation}
Multiplication by the corresponding magnetizations $m_i$ of states $S_i$ yields the
viscous evolution of remanence
\begin{equation}\label{VRMt}
    m(t) ~=~ \sum \limits_i m_i\, \rho_i(t).
\end{equation}
When a small field $H$ is applied, the energy barrier
$E_b^{ij}$ in first order  changes according to
\begin{equation}\label{field-barr}
    E_b^{ij}(H)~=~  E_b^{ij}+ (m_j - m_{ij}^{\max})\, H,
\end{equation}
where $m_{ij}^{\max}$ denotes the magnetization  at the  maximum
energy state along the optimal transition path from $S_i$ to $S_j$.
The approximation used to obtain (\ref{field-barr}) assumes that $H$ is so small
that it does not change the magnetization structures of the LEM and saddle-point states noticeably.
Only the field induced energy is taken into account.
It is easily seen that all other energy changes are of second order in $H$.

Using the in-field energy barriers it is straightforward  to determine the
matrix exponential which governs VRM acquisition. By defining
\begin{eqnarray}
% \nonumber to remove numbering (before each equation)
  \mu_{ij}(H) &=& -\frac{E_b^{ij}(H)}{k_B T\,\tau_0} ~~\mbox{for}~~i\neq j\\
  \mu_{ii}(H)&=& -\sum \limits_{i\neq j}  \mu_{ji}(H),
\end{eqnarray} the above zero-field theory automatically  extends to the weak field case.

In   case of our cubic PSD particle, all matrices are of size $60\times60$
and the calculations have been performed by a Mathematica (\copyright Wolfram Research)
program.

\subsection{Viscous remanence acquisition and decay in an ensemble of cubic PSD magnetite}
Using the above mathematical methods it is possible to calculate
the statistics of viscous remanence acquisition and decay for our
single PSD particle
with respect to any field vector of sufficiently small length $H$.
In order to model an isotropic ensemble, it is necessary to average the
VRM properties over all possible field directions. This has been approximated by drawing 20
random directions  from an equi-distribution over the unit sphere and averaging the
modelled VRM acquisition and decay curves.
For room temperature this yielded   the ensemble curve as shown in Fig.~\ref{ViscAcq}.
~\\[3mm]%
\par
\vbox{ \centerline{\hbox{ \psfig{figure=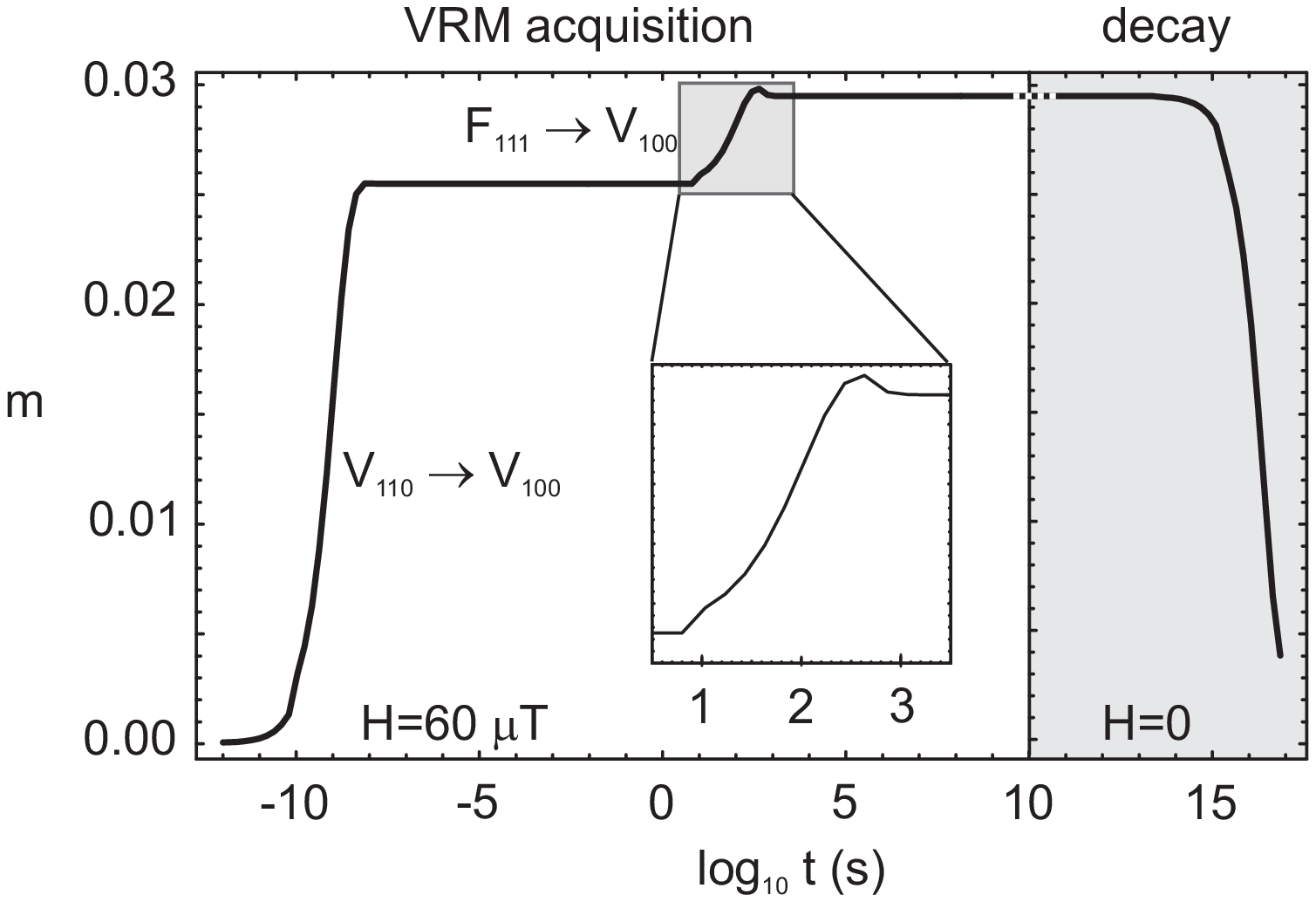,width=120mm}
}} \footnotesize
%  \begin{center}
{\bf Figure \bild{ViscAcq} :}
Modelled acquisition of viscous magnetization in the
cubic particle with $\lambda = 5.0$. The initial state is
an
equi-distribution over all possible LEM states with zero net magnetization.
In a
small external field the first acquisition process is the
immediate decay from V-$110$ type states into V-$100$ type states,
which occurs within about $10^{-9}$s.
Due to the field induced asymmetry of the energy barriers, a remanence is acquired
during this process. The second process is a decay of
F-$111$ type states into V-$100$ type states. This occurs between about $10^2$s and $10^3$s
and shows an intermediate overshooting of remanence.
%  \end{center}
\normalsize }
~\\[0.2cm]%
The left hand side of Fig.~\ref{ViscAcq} shows the remanence acquisition in a modelled
field of $H=60\mu$T when starting from an initial state $\rho_0$ at $t=0$ which assigns equal probability
to all existing LEM states.
Already within $10^{-9}$s the remanence increases rapidly due to the immediate
depletion of the nearly unstable V-$110$ vortex states which decay into the
  stable
V-$100$ states (see Table~\ref{trans-barr-2}).
The remanence forms because in zero field there are two equally probable
transitions, e.g. V-$110~\rightarrow$~V-$100$ and V-$110~\rightarrow$~V-$010$.
Within the external field one of these decay paths becomes more probable which leads to
a relative overpopulation of the field aligned V-$100$ type states.
Nearly synchronously there occurs a two step process
V-$111~\rightarrow$~V-$110~\rightarrow$~V-$100$.It
is controlled by the somewhat slower transition V-$111~\rightarrow$~V-$110$, but still
both take place within the first few $10^{-9}$s.
The last VRM acquisition processes occurs only after a much longer waiting time of $10-10^3$s.
First the initial F-$111$ type states transform via F-$110$ type states into V-$110$ type states
which then immediately decay into V-$100$
(Fig.~\ref{state-scheme}).
This last process  produces an astonishing remanence overshooting as displayed in
Fig.~\ref{ViscAcq}: The remanence during the VRM acquisition process is for a certain time
higher than the finally obtained equilibrium VRM.
In the next section we will show that is is not an artifact of the modelling,
but can be explained by a  real physical process.

The right hand side of  Fig.~\ref{ViscAcq} shows that
when the field is switched off after VRM acquisition, the obtained remanence is carried
only by   extremely stable  V-$100$-type states which require a   theoretical
waiting time of $10^{15}$s to equilibrate into a zero remanence state. 

%% file: discu.tex
\subsubsection{Viscous magnetization anomalies}
One of the most astonishing results of this study with respect to viscous magnetization is the predicted transient
increase of remanence during
the VRM acquisition. To understand this phenomenon more closely, we
give a physical explanation of this effect in terms of a simplified model.

The in field potential barrier is
\begin{equation}
E_s - E_F + (m_F - m_s) h.
\end{equation}
Here $E_s$ is the energy at the saddle point and $E_F$ is the zero field
energy of the flower state.
$W+$ and $W-$ are the number of grains (probability) in the vortex $V-100$
parallel and antiparallel to the external field $h$ in $x$-direction.

\subsubsection{High stability of PSD VRM}
Extremely stable VRM has been often observed in paleomagnetic studies.
The above mechanisms give a first theoretical explanation why
high stability of VRM should occur in PSD samples.
The basic process is the relaxation of naturally produced
metastable states into stable ones.

This typically occurs for TRM acquisition in PSD ensembles where the cooling rate is fast enough to
stabilize metastable flower states.
Then a long term VRM, acquired in the field after cooling and carried by newly formed
 vortex states is extremely
stable and can significantly bias any paleomagnetic measurement, especially paleointensity determinations.

Laboratory AF demagnetization  rather leads to a more stable
LEM state (perhaps even the GEM) because the magnetization structure is
provided with a lot of energy which is stepwise reduced.
Thus AF-demagnetization would rather end up in a vortex state for a PSD particle.

During natural chemo-viscous (VCRM) magnetization by crystal growth a grain
changes sequential from the  SP state into a stable SD and later a PSD state.
The first stable SD state is almost homogeneously magnetized along a $\langle 111\rangle$-axis.
It then transforms in a more developed flower state which then
becomes metastable as soon as the vortex has lower energy.
At this point,  the process of VRM acquisition will starts to
produce extremely stable remanences.

%% file: append.tex
\subsubsection*{Details of calculation}
This uses tensor calculus and Einstein's sum convention and for
comparison with classical physics interprets the variable $w$ as a time $t$ and $x$ as a
generalized variable $q$.
We have to apply the Euler-Lagrange operator
$\frac{\partial}{\partial q} -  \frac{d}{dt}\frac{\partial}{\partial \dot{q}}$
to the Lagrange function $L(q,\dot{q}) ~=~ \sqrt{g_i\,g_i } \,
 \sqrt{ \dot{q}_i \dot{q}_i}$, where $g_i ~=~ \partial_i E(q)$.
This needs the following expressions
\begin{equation}\label{NR1}
  \dot{\partial}_k \|\dot{q}\|~:=~\frac{\partial}{\partial \dot{q}_k}\, \sqrt{ \dot{q}_i \dot{q}_i}~=~\frac{\dot{q}_k}{\|\dot{q}\|}
\end{equation}
\begin{equation}\label{NR2}
 \frac{d}{dt} \,\dot{\partial}_k \|\dot{q}\| ~=~ \frac{d}{dt}\,\frac{\dot{q}_k}{\|\dot{q}\|}~=~
\frac{\ddot{q}_k}{\|\dot{q}\|}-\frac{\dot{q}_k\,(\dot{q}_j\,\ddot{q}_j)}{\|\dot{q}\|^3}~=~
\frac{\ddot{q}_k}{\|\dot{q}\|}.
\end{equation}
The last equation uses the fact that in the chosen parametrisation
 the tangent vector $\dot{q}$ has constant length along the path
and thus   is perpendicular to $\ddot{q}$.
Further,
\begin{equation}\label{NR3}
\partial_k (g_i\,g_i)^{1/2}~=~\frac{ g_j\,\partial_k g_j}{(g_i\,g_i)^{1/2}}~=~\frac{ g_j\,\partial_k g_j}{\|g\|}.
 \end{equation}
Putting this together results in
\begin{equation}\label{NR4}
\left( \frac{\partial}{\partial q} -  \frac{d}{dt}\frac{\partial}{\partial \dot{q}}\right)\,L(q,\dot{q})
 ~=~
 \|\dot{q}\|\,\partial_k\,\|g\| - \|g\|  \frac{d}{dt} \,\dot{\partial}_k \|\dot{q}\|~=~
 \frac{\|\dot{q}\|}{\|g\|}\, g_j\,\partial_k g_j - \frac{\|g\|}{\|\dot{q}\|}\,\ddot{q}_k.
  \end{equation}
Using the arc length parametrisation where $\|\dot{q}\|=1$,
this finally leads to the Euler-Lagrange equation
\begin{equation}\label{EL-3}
\ddot{q}~=~ \frac12 \,\frac{\nabla\, \|g\|^2}{\|g\|^2}~=~  \nabla\,\log \|g\|.
\end{equation}